\documentclass[10pt,journal,compsoc]{IEEEtran}

\usepackage{xcolor}
\usepackage{hyperref}

\ifCLASSOPTIONcompsoc
  \usepackage[nocompress]{cite}
\else
  \usepackage{cite}
\fi

\usepackage[numbers]{natbib}

\usepackage{subcaption}
\usepackage{graphicx}
\usepackage{multirow}
\usepackage{caption}
\newcommand\ExerciseCaption[1]{%
  \captionsetup{font=small}%
  \caption{#1}}
  
\usepackage{booktabs}

\ifCLASSINFOpdf

\else

\fi

\usepackage{multirow}
\usepackage{booktabs}
\usepackage{tabularx}

\usepackage{amsmath}

\begin{document}

\title{Identifying Similar Test Cases That \\Are Specified in Natural Language}

\author{Markos~Viggiato,
        Dale Paas,
        Chris Buzon,
        Cor-Paul~Bezemer
\IEEEcompsocitemizethanks{\IEEEcompsocthanksitem Markos Viggiato and Cor-Paul Bezemer are with the Analytics of Software, Games and Repository Data (ASGAARD) lab at the University of Alberta, Canada.
E-mail:\{viggiato, bezemer\}@ualberta.ca

\IEEEcompsocthanksitem Dale Paas and Chris Buzon are with Prodigy Education, Toronto, Canada.
E-mail: \{dale.paas, christopher.buzon\}@prodigygame.com
}
}

\markboth{}%
{Viggiato \MakeLowercase{\textit{et al.}}}
%

\IEEEtitleabstractindextext{%
\begin{abstract}
Software testing is still a manual process in many industries, despite the recent improvements in automated testing techniques. As a result, test cases are often specified in natural language by different employees and many redundant test cases might exist in the test suite. This increases the (already high) cost of test execution. Manually identifying similar test cases is a time-consuming and error-prone task. Therefore, in this paper, we propose an unsupervised approach to identify similar test cases. Our approach uses a combination of text embedding, text similarity and clustering techniques to identify similar test cases. We evaluate five different text embedding techniques, two text similarity metrics, and two clustering techniques to cluster similar test steps and four techniques to identify similar test cases from the test step clusters. Through an evaluation in an industrial setting, we showed that our approach achieves a high performance to cluster test steps (an F-score of 87.39\%) and identify similar test cases (an F-score of 83.47\%). Furthermore, a validation with developers indicates several different practical usages of our approach (such as identifying redundant and legacy test cases), which help to reduce the testing manual effort and time. 

\end{abstract}

\begin{IEEEkeywords}
Software testing, Test case similarity, Clustering.
\end{IEEEkeywords}}

\maketitle

\IEEEdisplaynontitleabstractindextext

\IEEEpeerreviewmaketitle

\newcommand{\rqone}[1]{RQ1: How effectively can we identify similar test steps that are written in natural language?}

\newcommand{\rqtwo}[1]{RQ2: How can we leverage clusters of test steps to identify similar test cases?}

\IEEEraisesectionheading{
\section{Introduction}\label{sec:introduction}
}

\IEEEPARstart{D}{espite} the many recent improvements in automated software testing, testing is still a manual process in many industries. For example, in the gaming industry, game developers face several challenges and difficulties with writing automated tests~\cite{politowski2021survey,pascarella2018video,murphy2014cowboys}. As a result, test cases are often described in natural language and consist of a sequence of steps that must be manually performed to test the target game. Furthermore, those test cases are usually defined by employees from different departments, such as Quality Assurance (QA) engineers or developers, which may result in redundant test cases (i.e., test cases that are semantically similar or even duplicates) as the system evolves and the test suite grows~\cite{rothermel2002empirical}. Having redundant test cases is problematic in particular in a manual testing scenario, due to the tediousness and cost of executing such manual tests.


Manually identifying similar or duplicate test cases to reduce test redundancy is an expensive and time-consuming task. In addition, naive approaches (e.g., searching for exactly matching test cases) are not sufficient to capture all similarity, as different test case writers may use different terminology to specify a test case, even for similar test objectives. Therefore, an automated technique to identify similar test cases is necessary as it can prevent the QA and development teams from wastefully executing test cases that perform the same task. Throughout this paper, for brevity we adopt the term ``similar test cases'' to refer to semantically similar and duplicate test cases.


In this paper, we propose an approach to identify similar test cases that are specified in natural language. More specifically, (1)~we use text embedding, text similarity, and clustering techniques to cluster similar test steps that compose test cases and (2)~we compare test cases based on their similarity in terms of steps that belong to the same cluster. 

In the first part of the study, we study how text embeddings obtained from different techniques, text similarity metrics, and different clustering algorithms can be leveraged to identify semantically similar test steps. We compare embeddings from five different techniques (Word2Vec, BERT, Sentence-BERT, Universal Sentence Encoder, and TF-IDF), two similarity metrics (Word Mover’s Distance and cosine similarity), and evaluate two different clustering techniques (Hierarchical Agglomerative Clustering and K-Means). In particular, we address the following research question for this part of the study:

\noindent\textbf{\rqone}\\
\textit{Understanding if we can effectively identify similar test steps automatically allows to know if we can rely on test step clusters to identify similarity between entire test cases. We found that we can achieve the highest performance (an F-score of 87.39\%) using an ensemble approach that consists of different embedding and clustering techniques. However, the best-performing single technique (Word2Vec) is very close in performance (with an F-score of 86.99\%).}

In the second part of the study, we leverage the previously detected clusters of test steps to identify similar test cases. We compared four different techniques to compute a similarity score (using the simple overlap, Jaccard, and cosine metrics) to measure the similarity of test cases based on the test step clusters that they have in common. In particular, we address the following research question for this part of the study:

\noindent\textbf{\rqtwo}\\
\textit{Given the difficulty of identifying similar test cases, which are usually composed of several steps, we use clusters of similar test steps to identify similar test cases. We found that test step clusters can be used to identify test case similarity with a high performance (an F-score of 83.47\%).}

Our work presents an approach to identify similar test cases based only on their natural language descriptions. We highlight that our approach is unsupervised as it does not require labelled data nor requires human supervision. In addition, no test source code or system model is necessary. QA engineers and developers can use our approach to obtain groups of similar test cases, which can be used to identify and remove redundant and legacy test cases from the test suite. Furthermore, existing groups of similar test cases can be leveraged to support the design of new test cases and help to maintain a more consistent and homogeneous terminology across the test suite. Finally, we provide access to the source code of our approach and the experiments that we performed.\footnote{\url{https://github.com/asgaardlab/test-case-similarity-technique}}

The remainder of the paper is organized as follows. In Section~\ref{sec:background}, we present the background information about text embedding and clustering techniques. We discuss related work in Section~\ref{sec:related_work} and our proposed approach in Section~\ref{sec:approach}. Section~\ref{sec:dataset} presents the dataset that we used to evaluate our approach. Sections~\ref{sec:approach_similar_steps} and~\ref{sec:approach_similar_cases} discuss the experiments that we performed to evaluate the two main stages of our approach. In Section~\ref{sec:discussions}, we discuss our results and the approach validation. Finally, Sections~\ref{sec:threats} and~\ref{sec:conclusion} present the threats to validity and conclude our work, respectively.

\section{Background}\label{sec:background}
In this section, we present an overview of the terminology and concepts that we use throughout the paper. In this work, we use ``test cases'' to refer to manual test cases that are described in natural language as a sequence of steps, i.e., test cases for which there is no source code associated.

\subsection{Text Representation}
In order to use text data as input for a machine learning algorithm, we first need to convert the text into a numeric vector through a process called \emph{text embedding}~\cite{weiss2010text,weiss2015fundamentals}. Different methods can be used to obtain a text embedding, and the embedding can be done at different granularity levels, such as at the word and sentence level. Below, we explain the different techniques that we use in this work to obtain the numeric representation of words and sentences.

\subsubsection{Word Embedding}
\label{sec:word_embedding}
\begin{figure*}[!t]
\centering
\subfloat[Examples of word embeddings for test steps.]{
	\label{fig:word_embedding_example}
	\includegraphics[width=0.7\textwidth]{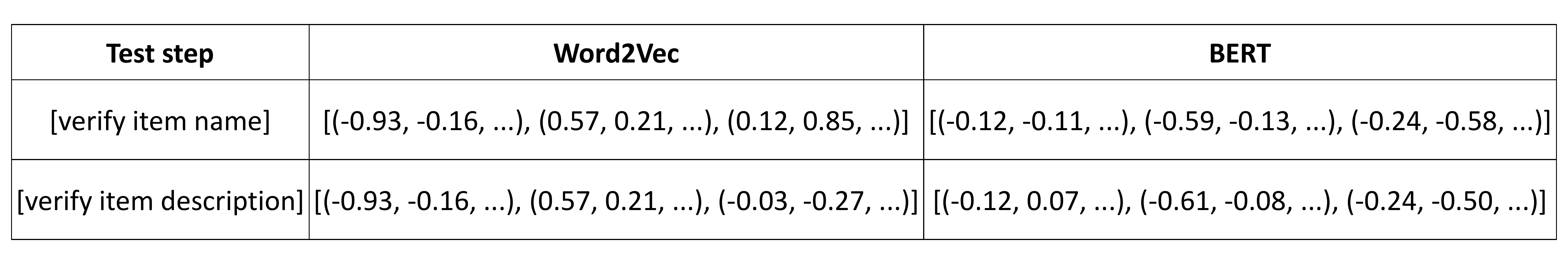} } 
 
\subfloat[Examples of sentence embeddings for test steps.]{
	\label{fig:sentence_embedding_example}
	\includegraphics[width=0.7\textwidth]{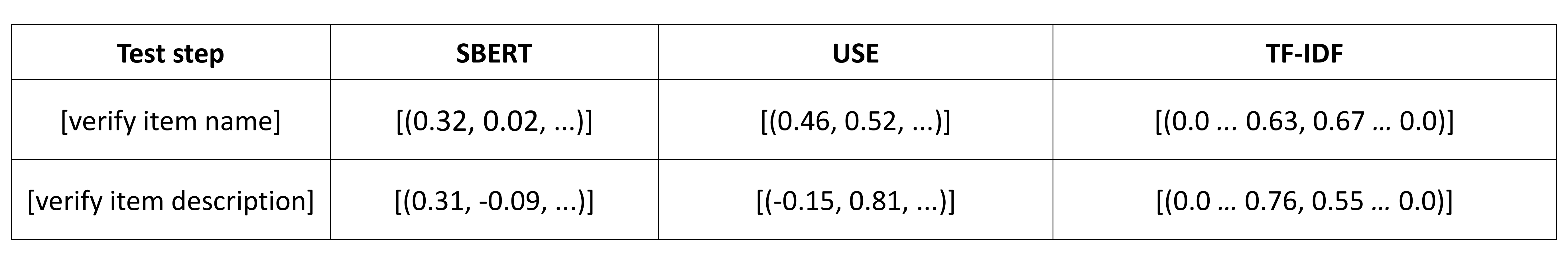} } 
 
\caption{Examples of test step embeddings. Note that we provide only the first two elements of the embedding vector due to space constraints as the actual vectors have a high dimension.}
\label{fig:text_embedding_examples}
\end{figure*}

A word embedding is the representation of a single word through a real-valued (and usually high-dimensional) numeric vector. In this study, we use two natural language processing techniques to obtain embeddings at the word level: Word2Vec~\cite{mikolov2013distributed} and BERT~\cite{DBLP:journals/corr/abs-1810-04805}. Figure~\ref{fig:word_embedding_example} presents two examples of pre-processed test steps along with part of their word embeddings obtained by the Word2Vec and BERT models. Next, we explain how each word embedding technique works and how the example embeddings presented in Figure~\ref{fig:word_embedding_example} are computed.


\textbf{Word2Vec} transforms words into high-dimensional numeric vectors that are able to maintain the syntactic and semantic relationships between words in the vector space~\cite{mikolov2013distributed,mikolov2013efficient}. This means that embeddings of similar words will (most of the time) be close in the vector space (i.e., the distance between the embedding vectors is small). Furthermore, with Word2Vec, each word is assigned a single numeric vector regardless of the context in which it is used, as we can see for the words ``verify'' and ``item'' in the two steps in Figure~\ref{fig:word_embedding_example}. In this work, we used the continuous bag-of-words (CBOW) model architecture of Word2Vec, which is faster than the other possible architecture, called skip-gram~\cite{mikolov2013efficient}. 

Differently from Word2Vec,~\textbf{BERT (Bidirectional Encoder Representations from Transformers)} is a transformer-based model that can be used to extract contextual word embeddings, i.e., embeddings that change depending on the context in which a word is present~\cite{DBLP:journals/corr/abs-1810-04805}. This means that the same word may have different embedding vectors, as we can see in Figure~\ref{fig:word_embedding_example}, where the BERT embeddings for the words ``verify'' and ``item'' are different in the two test steps because those words are in different contexts.

BERT is available as a model that was pre-trained on lower-cased English text (uncased BERT). This pre-trained model can further be trained with a domain-specific training set (known as domain-adaptive pre-training~\cite{gururangan2020don}). The BERT model uses WordPiece tokenization~\cite{wu2016google}, in which a word may be split into sub-words. For example, the word ``validate'' is composed of the sub-words ``valid'' and ``ate'', each one with its own embedding vector. Therefore, when extracting embeddings of words that are split into sub-words, we need to aggregate the embeddings of the sub-words (e.g., by averaging the embedding vectors).

\subsubsection{Sentence Embedding}
\label{sec:sentence_embedding}
Differently from word embedding, sentence embedding is the representation of a whole sentence with a real-valued (and usually high-dimensional) numeric vector. In this work, we use three different techniques to extract sentence embeddings (SBERT, USE, and TF-IDF). Figure~\ref{fig:sentence_embedding_example} presents two examples of pre-processed test steps along with part of their sentence embeddings obtained by the SBERT, USE, and TF-IDF techniques.
Next, we explain how each sentence embedding technique works.

\textbf{Sentence-BERT (SBERT)} is a BERT-based framework that allows us to directly extract numeric representations of full sentences~\cite{reimers2019sentence}. The embeddings of sentences that are semantically similar are close in the embedding space. We can use this information for different purposes, such as identifying paraphrases and clustering similar sentences. For instance, the SBERT embeddings of the two test steps presented in Figure~\ref{fig:sentence_embedding_example} are close in the embedding space (i.e., have a small distance between them). Among several generic and task-specific SBERT pre-trained models that are available\footnote{\href{https://www.sbert.net/docs/pretrained_models.html}{https://www.sbert.net/docs/pretrained$\_$models.html}}, three models are suitable for our task (identifying similar test steps): \textit{paraphrase-distilroberta-base-v1}, \textit{stsb-roberta-base}, and \textit{stsb-roberta-large}. While the first model is optimized to identify paraphrases and was trained on large scale paraphrase data, the second and third ones are the base and large versions of a model that was optimized for semantic textual similarity.





\textbf{Universal Sentence Encoder (USE)} is an embedding technique that can be used to directly extract embeddings from sentences, phrases, or short paragraphs to be used in another task, such as textual similarity and clustering tasks~\cite{cer2018universal}. With a similar behaviour to SBERT, the two examples presented in Figure~\ref{fig:sentence_embedding_example} have close embedding vectors.


Finally, we also used the \textbf{TF-IDF (Term Frequency–Inverse Document Frequency)} method to represent sentences. TF-IDF computes the importance of a word to a document by combining the word frequency in the document and the word frequency across all the other documents~\cite{joachims1996probabilistic,salton1991developments,jones1972statistical}. In our case, the test steps (i.e., sentences) are considered documents. We built a numeric vector for each test step using the word importance values. Words that are not present in the step are assigned a value of zero. We can observe a typical vector obtained with TF-IDF in the examples presented in Figure~\ref{fig:sentence_embedding_example}, in which the values different from zero correspond to the importance of the words presented in the ``verify item name'' and ``verify item description'' steps.

\subsection{Clustering techniques}

\textbf{Hierarchical Agglomerative Clustering (HAC)}~\cite{rokach2005clustering} is a clustering algorithm that works in a bottom-up manner. Initially, each data point corresponds to a single cluster itself, and as the algorithm iterates, different clusters are merged with the aim of minimizing a specific linkage criterion. The result of the iterative merging process is a tree structure that can represent the data points (and their clusters), known as a dendrogram. Although the dendrogram can be used to identify the number of clusters, in our work we determined that parameter empirically and used the number that maximizes our evaluation metric (as explained in Section~\ref{sec:clustering_test_steps}). Different linkage criteria can be used, such as single-linkage (the algorithm uses the minimum of the distances between all data points of two sets) and average-linkage (the algorithm uses the average of the distances between all data points of two sets).

The \textbf{K-means clustering}~\cite{duda1973pattern} algorithm splits the data points into $k$ different clusters. Different from HAC, no hierarchical cluster structure is generated with K-means. The goal of K-means is to group data points in order to minimize the distance between points belonging to the same cluster compared to the distance of points from different clusters. Using the Expectation-Maximization algorithm~\cite{moon1996expectation}, K-means starts with $k$ centroids. Then, the algorithm (1)~assigns each data point to the closest cluster (in terms of the distance between the point and the centroids) and (2)~computes the new centroids using the updated data point assignments. The execution finishes when there is no change to the allocation of data points.

\section{Related Work}\label{sec:related_work}
In this section, we discuss prior work that applied clustering techniques~\cite{li2020clustering,walter2018improving,chen2011using,zalmanovici2016cluster,b2016learning} and natural language processing (NLP)~\cite{masuda2016automatic,li2020clustering,wang2020automatic,wang2015automatic,mai2018natural,tan2020collaborative,masuda2015semantic} to software testing.

\subsection{Clustering techniques for software testing}


Our work is based on the study of~\citet{li2020clustering}, which proposed an approach to cluster test steps written in natural language based on the steps' similarities. The study used text embeddings (including embeddings obtained with the Word2Vec technique) together with the Relaxed Word Mover's Distance (RWMD) metric~\cite{kusner2015word} to measure similarity between embeddings. The test steps were then clustered with the hierarchical agglomerative and K-means clustering techniques. The approach was evaluated on a large-scale dataset of a mobile app and achieved an F-score of 81.55\% in the best case. The proposed approach also reduced the manual effort for implementing test-step API methods by 65.90\%. Differently from~\citeauthor{li2020clustering}'s work~\cite{li2020clustering}, we evaluated more recent NLP techniques to obtain word and sentence embeddings (BERT, SBERT, and Universal Sentence Encoder). Furthermore, we extended~\citeauthor{li2020clustering}'s work~\cite{li2020clustering} for the purpose of identifying similar test cases using the identified clusters of test steps.


\citet{walter2018improving} proposed an approach to improve the efficiency of test execution. The approach removes redundant test steps and uses clustering techniques to rearrange the remaining steps. To use the approach, the textual descriptions of test cases must be converted into a representation form of parameters concatenated by first order logic operators (AND, OR, NOT). The approach was evaluated in a case study with a system from an automotive industry company. The results indicated a test load reduction of 18\% due to the removal of redundant test steps and rearranging of the remaining steps.~\citet{chetouane2020using} proposed an approach to reduce a test suite by clustering similar test cases (based on their source code) with the K-means algorithm. 13 Java programs were used to evaluate if the approach could efficiently reduce the test suite and assess the impact on coverage metrics. The evaluation showed that the approach can reduce the test suite by 82.2\% while maintaining the same coverage metric as the original test suite.

\citet{pei2021dynamic} proposed distance-based Dynamic Random Testing (DRT) approaches with the goal of improving the fault detection effectiveness of DRT. The work clustered similar test cases based on their source code with three clustering methods: K-means, K-medoids, and hierarchical clustering. The information of distance between the test case groups was used to identify test cases that are closer to failure-causing groups. 12 versions of 4 open-source programs were used to evaluate the approaches. The evaluation showed that the proposed strategies achieve a larger fault detection effectiveness with a low computational cost compared to other DRT approaches.~\citet{arafeen2013test} investigated whether clustering of test cases based on similarities in their requirements could improve traditional test case prioritization techniques. The paper used TF-IDF and the K-means clustering algorithm to group test cases that have similar requirements. Two Java programs were used to evaluate the approach. The evaluation showed that the use of requirements similarity can improve the effectiveness of test case prioritization techniques but the improvements vary with the cluster size.

Differently from the works above, our study aims at finding redundant test cases by clustering similar test cases that are written in natural language. We experimented with different NLP and clustering techniques to find clusters of similar test steps, which are used with test case names to obtain similar test cases. Furthermore, differently from the work of~\citet{walter2018improving}, which converts natural language descriptions of test cases into a representation form of parameters concatenated by logic operators to be used with their approach, our proposed approach works in an unsupervised manner with the original test cases written in natural language.

\subsection{Natural Language Processing techniques for software testing}

\citet{wang2015automatic} proposed an approach to automate the generation of executable system test cases. The approach applies NLP techniques (such as tokenization and part-of-speech tagging) to textual data obtained from use case specifications. Furthermore, a domain model of the system under analysis is necessary to generate test data and oracles.~\citet{wang2015automatic} performed an industrial case study with automative software to demonstrate the feasibility of the proposed approach.~\citet{wang2020automatic} extended their previous work~\cite{wang2015automatic} by further providing empirical evidence about the scalability of the approach to generate executable, system-level test cases for acceptance testing from natural language requirements. In addition,~\citet{wang2020automatic} focused on embedded systems and demonstrated the effectiveness of the proposed approach using two industrial case studies, in which the approach correctly generated test cases that exercise different scenarios manually implemented by experts, including critical scenarios not previously considered.

\citet{yue2015rtcm} proposed a Test Case Specification (TCS) language, called Restricted Test Case Modeling (RTCM), and an automated test case generation tool, called \textit{aToucan4Test}, to transform textual test cases into executable test cases. RTCM provides a template that combines natural language with restriction rules and keywords for writing TCS. Two case studies were performed to assess the applicability of RTCM and a commercial video conferencing system was used to evaluate the \textit{aToucan4Test} tool. \textit{aToucan4Test} could correctly generate 246 executable test cases from 9 test case specifications of subsystems of the video conferencing system. The study also evaluated the effort to use RTCM and \textit{aToucan4Test} using the average time for deriving the executable test cases, which is 0.5 minutes.~\citet{mai2018natural} addressed the problem of automatically generating executable test cases from security requirements in natural language.~\citeauthor{mai2018natural} proposed an approach to generate security vulnerability test cases from use case specifications that capture malicious behavior of users. Similarly to previous work,~\citeauthor{mai2018natural} evaluated the approach with an industrial case study in the medical domain. The evaluation indicated that the proposed approach can automatically generate test cases detecting vulnerabilities. 

The aforementioned works used different NLP techniques to automatically generate different types of test cases. In contrast, we propose an approach that leverages different NLP techniques to extract text embeddings and can automatically identify similar test cases. The approach can be used to identify redundant and legacy test cases written in natural language.





\begin{table*}[!t]
\centering
\caption{Running example.}
\label{tab:test_case_running_example}
\begin{tabularx}{\linewidth}{p{1cm}p{2cm}p{1.5cm}p{1cm}XX} \toprule
\textbf{\begin{tabular}[c]{@{}l@{}}Test case\\ identifier\end{tabular}} & \textbf{\begin{tabular}[c]{@{}l@{}}Test case\\ name\end{tabular}} & \textbf{\begin{tabular}[c]{@{}l@{}}Test case\\ type\end{tabular}} & \textbf{\begin{tabular}[c]{@{}l@{}}Test step\\ identifier\end{tabular}} & \textbf{\begin{tabular}[c]{@{}l@{}}Test step\\ (before pre-processing)\end{tabular}}                                      & \textbf{\begin{tabular}[c]{@{}l@{}}Test step\\ (after pre-processing)\end{tabular}} \\ \midrule

\multirow{3}{*}{TC1}           & \multirow{3}{*}{\begin{tabular}[c]{@{}l@{}}Log in to an\\ existing account\end{tabular}}         & \multirow{3}{*}{Login}     & TS1.1                           & Login to the game using an existing account that has completed   the tutorial                                         & {[}login, game,   using, existing, account, completed, tutorial{]}                      \\
                                &                                                     &                            & TS1.2                           & Select the Playing from School portal                                                                                 & {[}select,   playing, school, portal{]}                                                 \\ \midrule
\multirow{7}{*}{TC2}          & \multirow{7}{*}{\begin{tabular}[c]{@{}l@{}}Assignment with\\ many students\end{tabular}}      & \multirow{7}{*}{Education} & TS2.1                         & Update the assignment adding students                                                                                 & {[}update,   assignment, adding, student{]}                                             \\
                                &                                                     &                            & TS2.2                         & Request the next skill and question from the algorithm gateway   for the 1st student on the assignment                & {[}request,   next, skill, question, algorithm, gateway, student, assignment{]}         \\
                                &                                                     &                            & TS2.3                         & Request the   next skill and question from the algorithm gateway for the middle student on   the assignment           & {[}request,   next, skill, question, algorithm, gateway, middle, student, assignment{]}
                                \\ \midrule

\multirow{10}{*}{TC3}          & \multirow{10}{*}{\begin{tabular}[c]{@{}l@{}}Student has\\ multiple\\ assignments\end{tabular}} & \multirow{10}{*}{Education} & TS3.1                         & Request the   next skill and question from the algorithm gateway for one of the students   that was on the assignment & {[}request,   next, skill, question, algorithm, gateway, one, student, assignment{]}    \\
                                &                                                     &                            & TS3.2                         & Remove student   from the first assignment                                                                            & {[}remove,   student, first, assignment{]}                                              \\
                                &                                                     &                            & TS3.3                         & Request the   next skill and question from the algorithm gateway for one of the students   that was on the assignment & {[}request,   next, skill, question, algorithm, gateway, one, student, assignment{]}    \\
                                &                                                     &                            & TS3.4                         & Remove the   student from the second assignment                                                                       & {[}remove,   student, second, assignment{]}                                
                                 \\ \bottomrule
\end{tabularx}
\end{table*}




\section{Proposed approach}\label{sec:approach}

\begin{figure}[!t]
\centering
\includegraphics[width=0.5\textwidth]{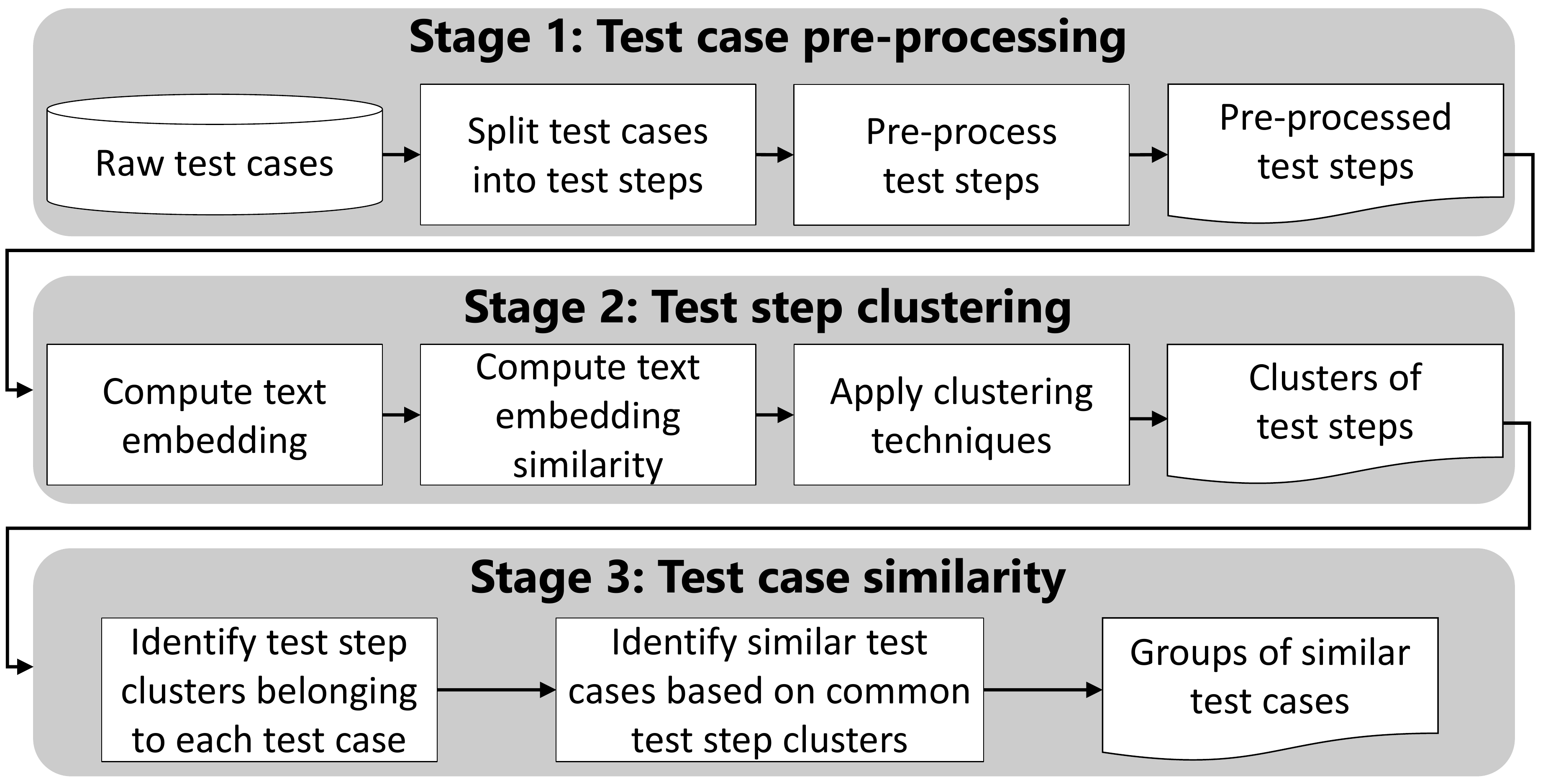}
\caption{Overview of our proposed approach.}
\label{fig:approach}
\end{figure}

In this section, we demonstrate our proposed approach through a running example. Figure~\ref{fig:approach} presents an overview of our approach, which consists of three stages: (1) pre-processing of test cases, (2) clustering of similar test steps and (3) identification of similar test cases. Next, we explain the stages of our approach.

\subsection{Stage 1: Test case pre-processing}
\label{sec:approach_preprocessing}
Our approach relies only on test cases that are written in natural language, which means that there is no source code available for our test cases. The input to our approach consists of unprocessed (raw) test cases. Table~\ref{tab:test_case_running_example} presents three test cases (TC1, TC2, and TC3) that we use as a running example to describe how our approach identifies similar test cases. As we can observe, each test case contains an identifier, a name and a type. In addition, a test case has one or more test steps, which are instructions that the tester must perform in order to achieve the overall objective of the test case. Note that this objective is generally not explicitly specified. The test steps that we collect to perform our experiments are explicitly identified (i.e., each test step has its own field within a test case). Therefore, we can directly collect the test steps and identify to which test case they belong.
Each test step is assigned a unique identifier and is pre-processed. Initially, we used tokenization to transform the step sentences into a list of words. To ensure that we have high-quality data, we obtained a list of the unique words in our data and manually inspected the list to identify misspelled words, which were used to build a list of [misspelled\_word, fixed\_word] tuples. The manually built tuple list was used to programmatically replace misspelled words with the corresponding fixed words across the entire dataset. We then removed stopwords (such as ``a'', ``of'', and ``the'') as they do not add meaning to the sentences. Also, we applied lemmatization to the words to have a consistent terminology across the data. Finally, similar to prior work~\cite{li2020clustering}, we removed words that occur only once in the whole dataset as they may introduce bias in the data (e.g., due to an imprecise numeric representation of those words). Overall, a test case instance can be represented by the triple: \[\textless test\_case\_name, test\_case\_type, test\_steps\textgreater\]

\vspace{-0.3cm}

\begin{figure*}[!t]
\centering
\includegraphics[width=0.80\textwidth]{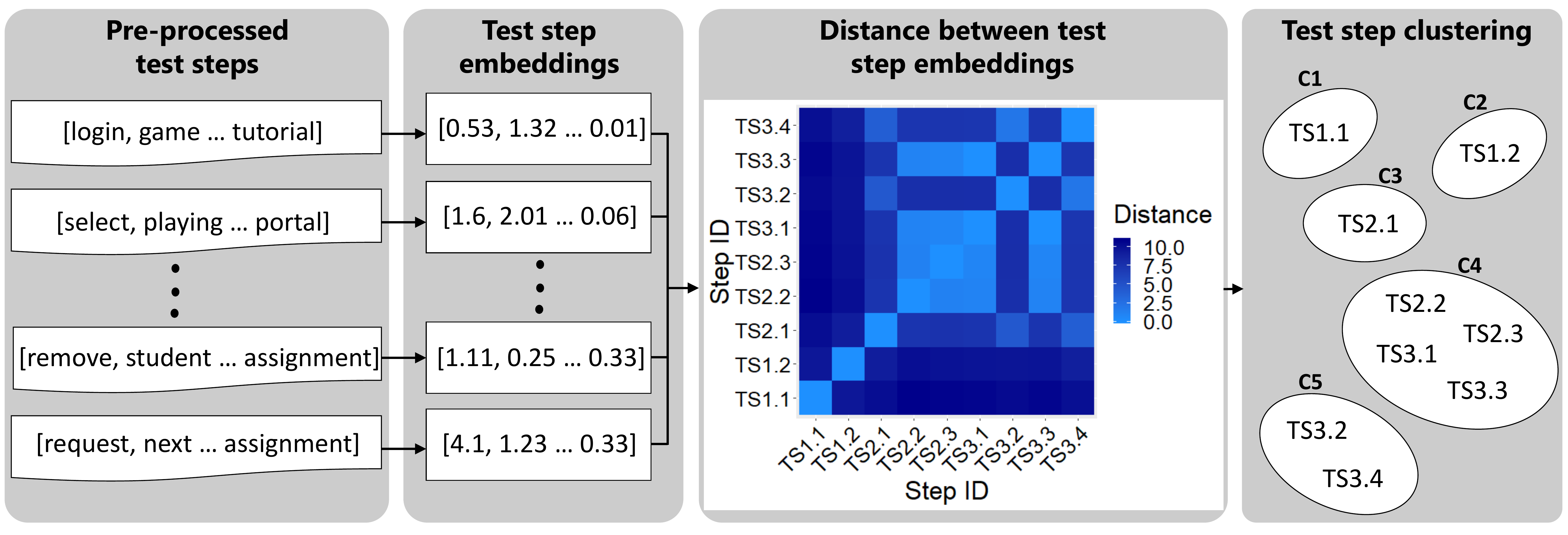}
\caption{Overview of stage 2 of our approach with the running example.}
\label{fig:approach_stage_2}
\end{figure*}

\begin{figure*}[!t]
\centering
\includegraphics[width=0.80\textwidth]{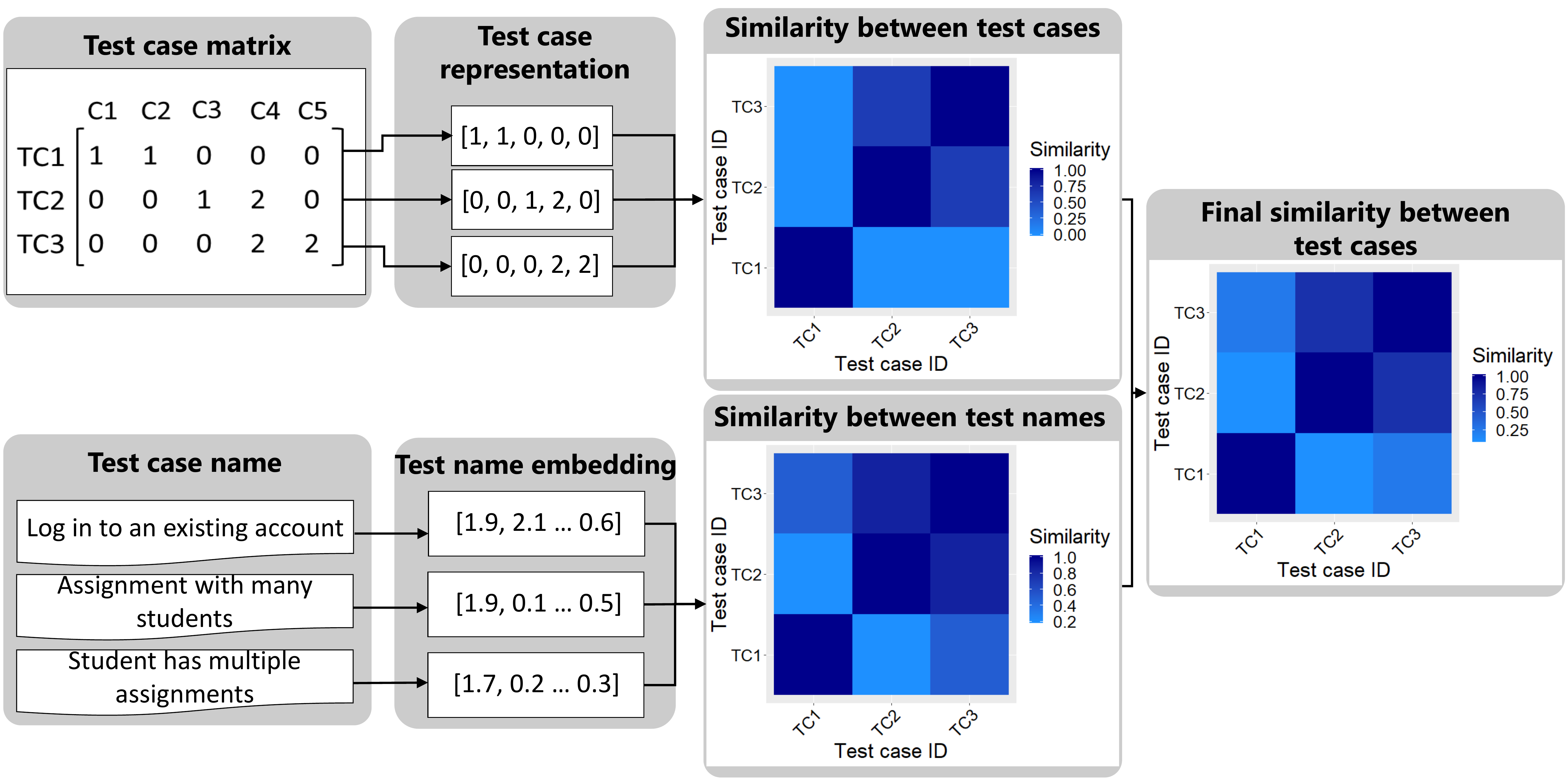}
\caption{Overview of stage 3 of our approach with the running example.}
\label{fig:approach_stage_3}
\end{figure*}

\subsection{Stage 2: Test step clustering}
In the second stage, our approach clusters similar test steps. Figure~\ref{fig:approach_stage_2} shows how the steps of the three test cases are processed in this stage. Before applying a machine learning algorithm to text data, we need to transform the text into a numeric representation~\cite{weiss2010text,weiss2015fundamentals}. Our approach starts by transforming each test step into one or more numeric vectors (text embedding). The pairwise similarity between steps (in terms of embedding distance) is then computed. The computed distances between the text embeddings of the test steps can be used to capture their similarity. In particular, embeddings that are close in the embedding space should represent similar steps.

Finally, our approach leverages the computed distances to identify clusters of similar test steps. While steps that have a small distance between them should belong to the same cluster, steps with larger distances should be in different clusters.



\subsection{Stage 3: Test case similarity}
In the last stage, our approach leverages the clusters of test steps identified in stage 2 together with the test case name to find similar test cases. Figure~\ref{fig:approach_stage_3} shows how the TC1, TC2, and TC3 test cases are processed in this stage. The relationship between test cases and test step clusters is represented through a matrix in which each row is a test case (TC1, TC2, and TC3) and each column is a step cluster (C1, C2, C3, C4, and C5). Initially, for each test case (matrix row), the approach identifies the test step clusters (matrix column(s)) that contain one or more steps of the test case. Our approach supports the use of Boolean  (which yield a matrix consisting of 0's and 1's) or numeric flags. Note that a numeric flag represents the number of test steps present in the identified cluster. After filling in the matrix, each test case is represented by the corresponding Boolean or numeric vector (a row in the matrix) with a length corresponding to the total number of test step clusters. Test cases are then compared to each other in terms of the similarity between their representation vectors. Finally, to incorporate knowledge from the test case name, the approach computes the pairwise similarity between test case name embeddings and combines this similarity metric with the one obtained from the test step clusters. The final test case similarity score is a weighted average between the test step cluster and the test case name metrics. For the running example, our approach identifies the TC2 and TC3 test cases as similar but both are different from the TC1 test case. A QA engineer can then investigate those test cases to decide, for example, whether they are redundant or should be improved.


\section{Dataset}\label{sec:dataset}
We collected 3,323 test case descriptions written in natural language. The test cases under study were manually designed to test the \textit{Prodigy Game}\footnote{https://www.prodigygame.com/main-en/}, a proprietary, educational math game with more than 100 million users around the world. Each test case is composed of one or more test steps and, in total, there are 15,644 steps. We also collected the predefined type of the test case regarding the part of the game that is being tested. The test case type is available for 2,053 test cases (62\% of the total number of test cases). All the test steps are pre-processed according to the pre-processing steps as explained in Section~\ref{sec:approach_preprocessing}. 


To evaluate the performance of our approach (stage 2, for test step clustering, and stage 3, for test case similarity), we used our dataset to manually build a ground truth of similar test steps (stage 2) and similar test cases (stage 3), as we explain below. 

\noindent\textbf{Ground truth of similar test steps (stage 2 of our approach).}
We selected a representative sample from all 15,644 test steps with a confidence level of 95\% and a confidence interval of 5\%, which corresponds to 394 steps. The test step samples were manually analyzed in an incremental manner: for each step under analysis, we looked at all the previously analyzed steps to identify the similar ones. When we found one step or a group of steps similar to the current step, we included the current one in the group of the previous steps. If there was no previous step similar to the current one, we assigned it to a new cluster. The ground truth of similar test steps ended up with a total of 213 clusters and an average of 1.9 (standard deviation of 2.0) test steps per cluster. We also found that the largest cluster has 15 test steps.

\noindent\textbf{Ground truth of similar test cases (stage 3 of our approach).}
We selected a representative sample of test cases with a confidence level of 95\% and a confidence interval of 5\%, which corresponds to 381 test cases. Similarly to the way that we built the ground truth of similar test steps, the test case samples were manually analyzed in an incremental manner: for each test case under analysis, we looked at all the previously analyzed cases to identify the similar ones in terms of what is being tested, the similarity between the test steps of the test cases, and similarity between the test case names. Note that, for each test case, we analyzed the test case name, test case type, and all the steps that compose the test case. The ground truth of similar test cases ended up with a total of 242 clusters and an average of 1.6 (standard deviation of 1.9) test cases per cluster. For this ground truth, we found that the largest cluster has 21 test cases.

\section{Evaluating our approach for clustering similar test steps}\label{sec:approach_similar_steps}
In this section, we discuss the experiments that we performed to evaluate our approach for clustering similar test steps in an industrial setting.


\subsection{Evaluated techniques}
\begin{figure}[!t]
\centering
\includegraphics[width=0.5\textwidth]{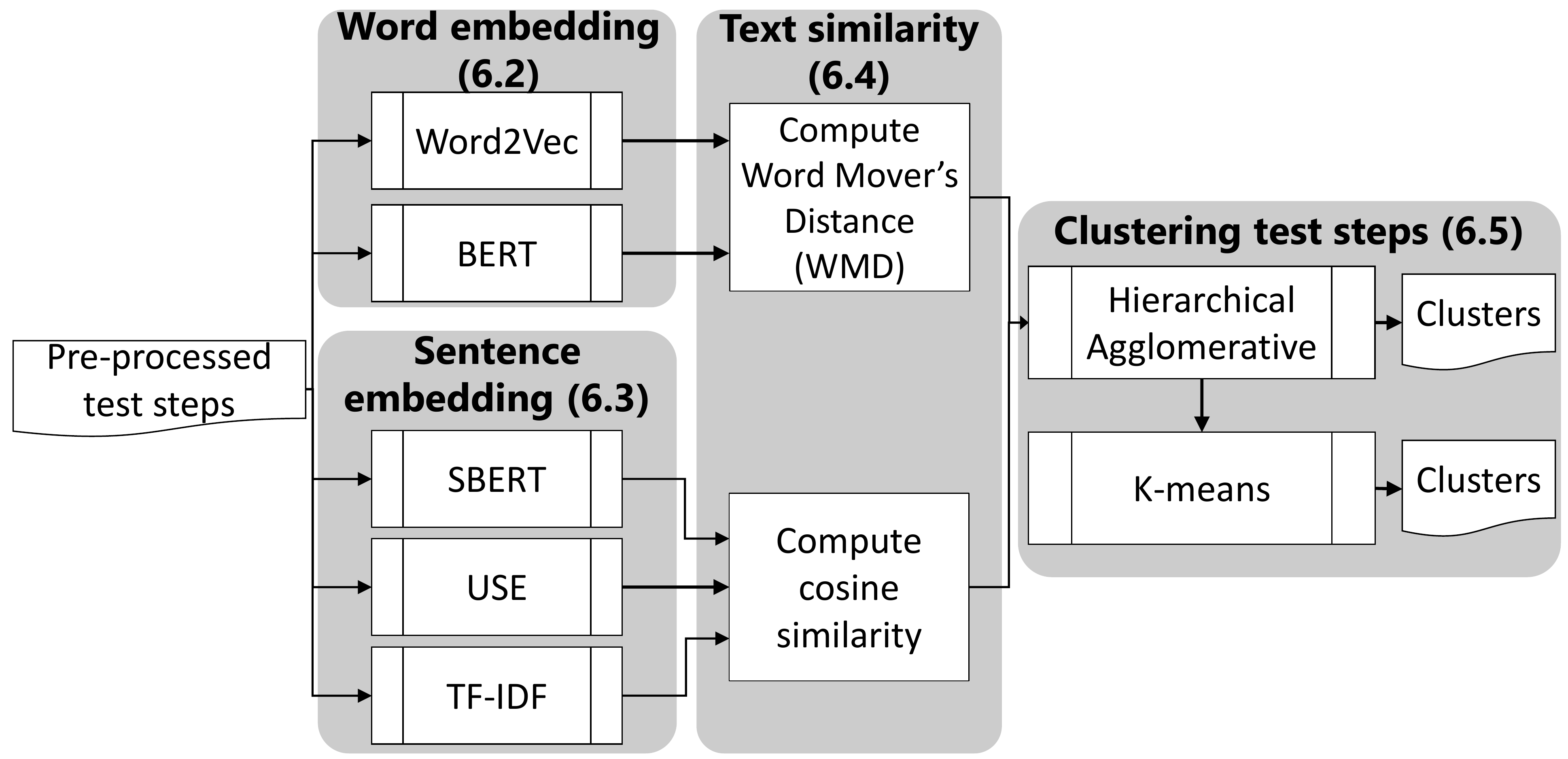}
\caption{Overview of the experiments to identify clusters of similar test steps.}
\label{fig:test_step_similarity_experiments}
\end{figure}

Our approach consists of several steps that can be implemented through different techniques and models. To evaluate our approach, we performed experiments with combinations of five different text embedding techniques, two similarity metrics, and two clustering techniques. Figure~\ref{fig:test_step_similarity_experiments} presents an overview of the experiments that we performed to address RQ1. Different NLP techniques can be used for text embedding at different granularities, such as words, sentences, and short paragraphs~\cite{mikolov2013distributed,mikolov2013efficient,DBLP:journals/corr/abs-1810-04805,reimers2019sentence,cer2018universal,le2014distributed}. As our test steps usually consist of a single sentence and the test steps are transformed into a list of words after pre-processing, we adopt word-level and sentence-level text embedding. We used two techniques to obtain text embeddings at the word-level (Word2Vec~\cite{mikolov2013distributed,mikolov2013efficient} and BERT~\cite{DBLP:journals/corr/abs-1810-04805}) for the test steps and computed the embedding similarity using the Word Mover's Distance (WMD) metric~\cite{kusner2015word}. For text embeddings at the sentence-level, we used three techniques (SBERT~\cite{reimers2019sentence}, Universal Sentence Encoder~\cite{cer2018universal}, and TF-IDF~\cite{joachims1996probabilistic,salton1991developments}) and used the cosine similarity to compare the embeddings. For both types of embeddings, we applied the hierarchical agglomerative~\cite{sneath1973numerical} and K-means~\cite{duda1973pattern} clustering techniques to obtain clusters of similar steps.


\subsection{Configuration of the word embedding techniques}
\label{sec:word_embedding_evaluation}

\noindent\textit{Word2Vec.}
We trained a Word2Vec model using all 15,644 test steps that we collected. Furthermore, to provide more context to the embedding model during training, we concatenated the test case type (available for 2,053 test cases) and test case name to each step. We used an embedding vector of length 300 (as in the original study that proposed the Word2Vec model~\cite{mikolov2013distributed}). We used the continuous bag-of-words (CBOW) model architecture of Word2Vec with two context words as this configuration provides the highest test step clustering performance, which was determined through an experiment in which we varied the number of context words from one to ten. We initialized the word embeddings with the weights from the large-scale pre-trained model released by Google.\footnote{\href{https://code.google.com/archive/p/word2vec/}{https://code.google.com/archive/p/word2vec/}} This model contains 3 million word embeddings with dimension 300 and was trained on a Google News corpus with approximately 100 billion words. For words that are present in our dataset but not in the pre-trained model (and, therefore, cannot be initialized with pre-trained weights), we followed a process proposed by~\citet{li2020clustering} to initialize the word embeddings. We computed the mean and standard deviation of the initialized words and initialized the remaining words with samples of a normal distribution parameterized by the computed mean and standard deviation. Finally, the outcome of the training process is the word embeddings learned with our data.

\vspace{0.15cm}
\noindent\textit{BERT.}
In this work, we used the pre-trained model released by Google\footnote{\href{https://github.com/google-research/bert}{https://github.com/google-research/bert}} (\emph{pre-trained BERT}) to obtain contextual embeddings of the test steps. Furthermore, we used a model with additional pre-training using our own corpus of test steps (\emph{domain-adaptive pre-trained BERT}) to obtain the contextual embeddings. We explain the configurations of both models below.

\noindent\underline{Pre-trained BERT.}
For the pre-trained model, we used the uncased (case-insensitive) version of the base model~\cite{DBLP:journals/corr/abs-1810-04805,turc2019well}. We transformed the test step text into the BERT format by adding the \textit{[CLS]} and \textit{[SEP]} tokens respectively to the start and end of each test step text. 
The test step was then tokenized with BERT's own tokenizer. Finally, we used the tokenized steps to extract the contextual embeddings. As explained in Section~\ref{sec:word_embedding}, we can adopt different pooling strategies to obtain the embedding vector for a word. We performed experiments with four different pooling strategies to combine the layers (as suggested by the original paper's authors~\cite{DBLP:journals/corr/abs-1810-04805}): using only the second-to-last layer, summing the last four layers, averaging the last four layers, and concatenating the last four layers. We found that summing the last four layers achieves the best performance with our data. Finally, we used the average of sub-word embeddings (see Section~\ref{sec:word_embedding}) to obtain the original out-of-vocabulary word embedding.

\noindent\underline{Domain-adaptive pre-trained BERT.}
We also performed additional pre-training of BERT with our corpus. For the additional pre-training, after experimenting with the base and large models, we decided to use the uncased version of the BERT large model as the initial checkpoint (i.e., we performed the additional pre-training on top of the pre-trained large model). We followed the same process to configure the test step text to a BERT-friendly format. However, differently from the pre-trained model, using the second-to-last layer (instead of summing the last four layers) achieves the best results for the domain-adaptive pre-trained BERT model.

\subsection{Configuration of the sentence embedding models}
\label{sec:sentence_embedding_evaluation}

\noindent\textit{Sentence-BERT (SBERT).}
We performed experiments with three available pre-trained SBERT models suitable for our task (see Section~\ref{sec:sentence_embedding}): \textit{paraphrase-distilroberta-base-v1}, \textit{stsb-roberta-base}, and \textit{stsb-roberta-large}. We decided to use the \textit{paraphrase-distilroberta-base-v1} model since it achieves the best results with our data. To obtain the embeddings for the test steps, we just provided the test steps directly as parameters to the SBERT model. 

\vspace{0.15cm}
\noindent\textit{Universal Sentence Encoder (USE).}
To obtain the test step embeddings with the USE model, we provided the steps directly as parameters to the USE model. 

\vspace{0.15cm}
\noindent\textit{TF-IDF.}
Finally, we also used TF-IDF to extract the numeric vector representations of the test steps. For each word, we computed its importance in a single test step relative to all the other test steps. 

\subsection{Computing the test step similarity}
\label{sec:computing_step_similarity}

\noindent\textit{Word Mover's Distance (WMD).}
We used the Word Mover's Distance (WMD)~\cite{kusner2015word} metric to measure the similarity between test step embeddings at the word level. The WMD metric is suitable to be used together with the Word2Vec and BERT models because of the property that distances between embedded words in the embedding space are semantically meaningful, which is a property that WMD relies on~\cite{kusner2015word}. We computed the pairwise WMD metric between any two test steps and built a distance matrix of dimension \textit{[15,644 x 15,644]}. The more similar two steps are, the lower is the WMD metric, with the lowest bound being zero for exactly matching steps. 

\vspace{0.15cm}
\noindent\textit{Cosine similarity.}
Since cosine is a widely used metric to measure similarity between text vectors~\cite{huang2008similarity,li2013distance,gomaa2013survey,salton1988term}, we used the cosine to measure the similarity between test step embeddings at the sentence level. Similarly to the way we computed the WMD metric, we computed the pairwise cosine similarity between any two test steps and built a distance matrix of dimension \textit{[15,644 x 15,644]}. As the cosine similarity value measure the cosine of the angle between two step numeric vectors, the smaller the angle, the larger its cosine and the more similar the two test steps are.

\subsection{Clustering test steps}
\label{sec:clustering_test_steps}

\noindent\textit{Hierarchical Agglomerative Clustering.}
We applied the hierarchical agglomerative clustering technique to the distance matrix that we built in the previous step (Section~\ref{sec:computing_step_similarity}). We used the average linkage criterion (with Euclidean distance), which means that the clustering algorithm merges pairs of test step clusters that minimize the average distance between each observation of the pairs.

\vspace{0.15cm}
\noindent\textit{K-means.}
To apply the K-means clustering technique, we used the test step embeddings obtained with the word/sentence embedding techniques (Sections~\ref{sec:word_embedding_evaluation} and~\ref{sec:sentence_embedding_evaluation}). Note that, for word-level embeddings, we transformed the embedding vectors of the words of a test step into a single vector to represent the whole test step by computing the word embeddings' average. Furthermore, to speed up the execution of K-means, we used the centroids of the clusters obtained by the hierarchical approach as the initialization centroids, similarly to prior work~\cite{li2020clustering,lu2008hierarchical}.


Regarding the number of clusters for both clustering techniques, we chose the number of clusters that maximized the F-score (which is our evaluation metric, as explained in Section~\ref{sec:evaluation_metric}). We performed a search by varying the number of clusters from 50 up to 15,000 with a step of 50, and for each value we executed both clustering approaches and computed the F-score. Finally, we selected the (optimal) number of clusters for which each clustering technique achieved the highest F-score. Note that the optimal number of clusters might be different for the hierarchical clustering and K-means.

\vspace{0.15cm}
\noindent\textbf{Ensemble approach.}
Each text embedding technique that we used has different characteristics and properties to extract word or sentence embeddings, which leads to different clusters of test steps. Therefore, attempting to mitigate each model's specific weaknesses, we built an ensemble approach that uses majority voting. That is, the ensemble approach assigns two test steps to the same cluster if at least three (out of the five) previously implemented approaches  assign those two steps to the same cluster.

\vspace{0.15cm}
\noindent\textbf{Baseline.}
We used two baselines to evaluate the performance of our proposed approaches. The first baseline assigns test steps to the same cluster only if those steps are exactly the same. The second baseline uses the Word2Vec technique together with the WMD similarity metric and only assigns two test steps to the same cluster if the WMD similarity of those steps is zero (i.e., their embeddings are the same).

\subsection{Evaluation metric}
\label{sec:evaluation_metric}
We are interested in penalizing both the false positives (to avoid excessive suggestions of similar test steps when they are not similar) and false negatives (to avoid missing out many similar test steps). Therefore, we used the F-score metric (as shown in Equation~\ref{eq:f_score}) to evaluate the test step clustering approaches as this metric captures the trade-off between precision (related to false positives) and recall (related to false negatives). Using the test steps present in the manually built ground truth of similar test steps, we analyzed all the pairs of test steps, similarly to prior work~\cite{li2020clustering}:
\begin{itemize}
    \item\textit{True positive (TP)}: when a pair of steps is included in the same cluster by our approach and the steps indeed belong to the same cluster in the ground truth.
    \item \textit{False positive (FP)}: when a pair of steps is included in the same cluster by our approach but the steps do not belong to the same cluster in the ground truth.
    \item \textit{True negative (TN)}: when a pair of steps is not included in the same cluster by our approach and the steps do not belong to the same cluster in the ground truth.
    \item \textit{False negative (FN)}: when a pair of steps is not included in the same cluster by our approach but the steps belong to the same cluster in the ground truth.
    
\end{itemize}

We then computed the F-score metric as follows:

\begin{equation}
\label{eq:f_score}
\text{F-score} = 2 \times \frac{precision \times recall}{precision + recall }
\end{equation}

Where the precision corresponds to the proportion of true positives regarding all the pairs identified as positive ($\frac{TP}{TP + FP}$) and the recall corresponds to the proportion of true positives regarding all the existing positive instance ($\frac{TP}{TP + FN}$).

\subsection{Findings}
\textbf{Similar test steps that are written in natural language can be identified with a high performance by applying the ensemble approach that uses majority voting. Furthermore, we can achieve a similar high performance using a single technique (Word2Vec).} Table~\ref{tab:f_score_approaches} presents the F-score of all the approaches that we evaluated along with the clustering techniques and the optimal number of clusters.


All the proposed approaches achieve a similar and high performance, with an F-score between 83.96\% and 87.39\%, except for both baselines, which have the same F-score of 70.40\%. More specifically, the ensemble approach achieves the highest performance, with an F-score of 87.39\%. If we look at the performance of the single models, Word2Vec with K-means has the highest F-score (86.99\%), which is very close to the ensemble approach performance. TF-IDF achieves the second highest F-score (86.67\%) among the single models, followed closely by SBERT with K-means (86.10\%) and Domain-BERT with K-means (85.77\%)

Regarding the two versions of the BERT model, we observe that the domain-adaptive pre-trained BERT is a little better, with F-scores of 85.60\% (using HAC) and 85.77\% (using K-means), in comparison to the generic pre-trained BERT, with F-scores of 84.65\% (using HAC) and 85.15\% (K-means). One possible reason for the small gain is that we do not have large amounts of data for the domain-adaptive pre-training. However, our findings indicate that the additional pre-training is capable of improving the model performance and might be more helpful with larger datasets.

We can observe that for all the approaches except for TF-IDF, running K-means on top of HAC is beneficial as this increases the F-score. Note, however, that the gain in performance is minimal, such as 1.14\% and 0.50\% in absolute percentage point for Word2Vec and BERT, respectively. On average, applying K-means on top of hierarchical clustering increases the performance by 0.33\% in absolute percentage point.

Finally, even though the ensemble approach has the highest performance for clustering test steps, this approach is computationally expensive as it requires the implementation and execution of all the other approaches, which takes around 6 hours in total using a single core on an Intel i7-8700 CPU. However, we can achieve a very close performance with a single technique, such as Word2Vec (which takes around 2.5 hours to execute), TF-IDF (which takes around 25 minutes), or SBERT (which takes around 3 minutes). Our experiments showed that both Word2Vec and BERT present a large execution time due to the large computational cost of computing the Word Mover's Distance. The reported execution times are for the full test step clustering pipeline (test step pre-processing, word embedding training, test step similarity, and clustering).



\begin{table}[!t]
\centering
\caption{F-score of the test step clustering approaches along with the clustering techniques and the optimal number of clusters obtained (\# Clusters). Note that we abbreviate hierarchical agglomerative clustering as HAC. The best performing approaches are highlighted in bold.}
\label{tab:f_score_approaches}
\begin{tabular}{lp{2.099cm}ll} \toprule
Text embedding technique             & Clustering & F-score & \# Clusters \\ \midrule
Baseline 1           & -                   & 70.40    & 4407                \\
Baseline 2           & -                   & 70.40    & 4393                \\
Word2Vec             & HAC                  & 85.85   & 2650                \\
\textbf{Word2Vec}   & \textbf{K-means}    & \textbf{86.99}   & \textbf{2650}              \\
BERT                 & HAC                  & 84.65   & 3050                \\
BERT                 & K-means        & 85.15   & 3050                \\
Domain-adaptive BERT & HAC                  & 85.60    & 3300                \\
Domain-adaptive BERT & K-means        & 85.77   & 3300                \\
SBERT                & HAC                  & 85.71   & 3350                \\
SBERT                & K-means        & 86.10    & 3350                \\
USE                  & HAC                  & 83.96   & 3050                \\
USE                  & K-means        & 84.39   & 2900                \\
TF-IDF               & HAC                  & 86.67   & 2500                \\
TF-IDF               & K-means        & 86.03   & 2500                \\
\textbf{Ensemble approach}    & -           & \textbf{87.39}   & \textbf{3158}               \\ \bottomrule
\end{tabular}
\end{table}


\section{Evaluating our approach for identifying similar test cases}\label{sec:approach_similar_cases}
In this section, we discuss the experiments that we performed to evaluate our approach for identifying similar test cases that are specified in natural language. Below, we discuss four different techniques to identify similar test cases using the previously identified clusters of test steps.


\subsection{Evaluated techniques}
\begin{figure}[!h]
\centering
\includegraphics[width=0.5\textwidth]{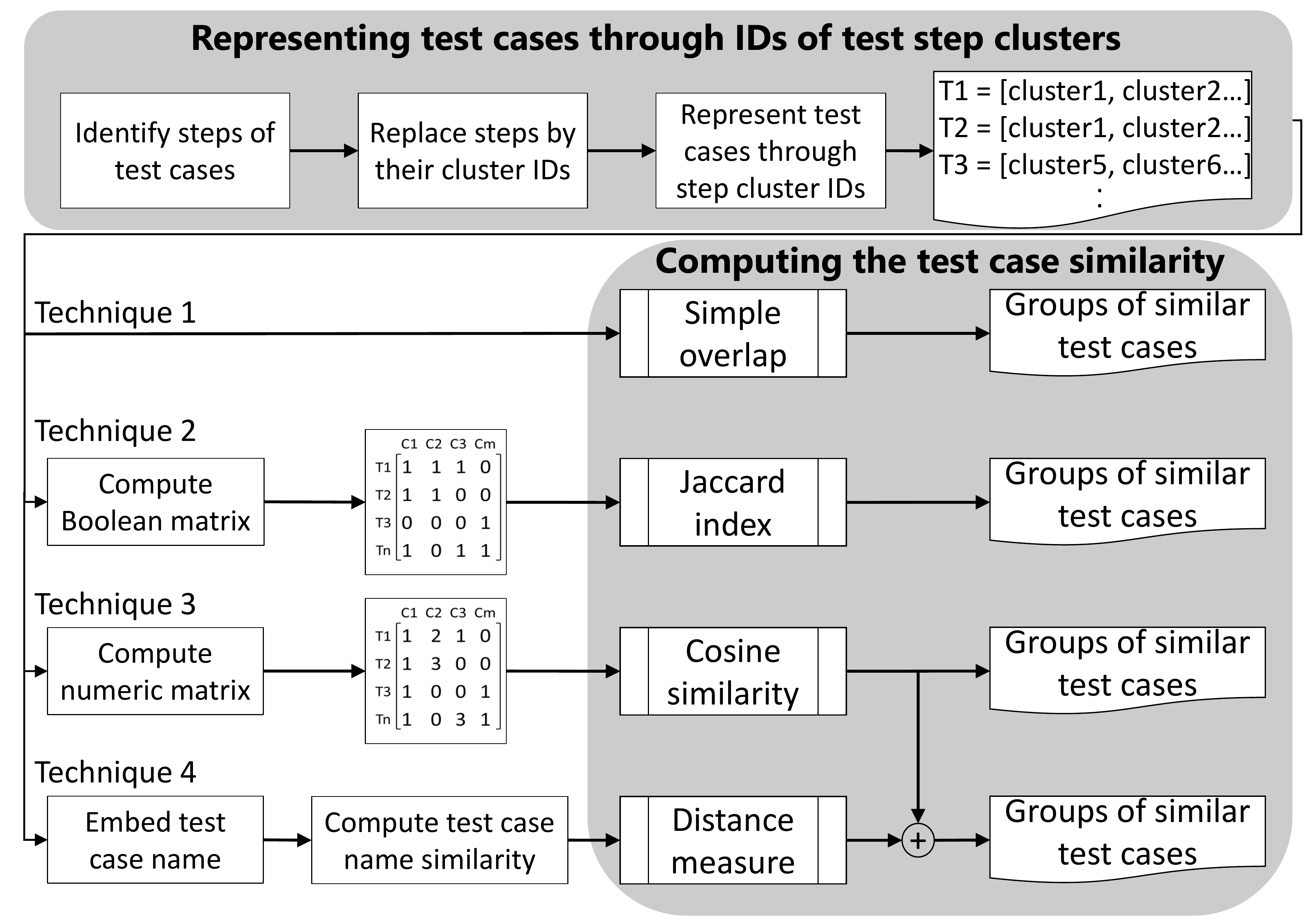}
\caption{Overview of the experiments to identify similar test cases.}
\label{fig:test_case_similarity}
\end{figure}

We performed experiments with four different proposed techniques to identify similar test cases using the previously identified clusters of test steps. Figure~\ref{fig:test_case_similarity} gives an overview of the experiments. To explain how each technique works, we use the two example test cases presented in Table~\ref{tab:methods_case_similaririty}.


\begin{table*}[!t]
\centering
\caption{Examples of test case representations (through vectors) obtained with the experimented four techniques and the corresponding similarity scores.}
\label{tab:methods_case_similaririty}
\begin{tabular}{lllllll} \toprule
\textbf{Test case} & \textbf{Test step}        & \textbf{Test step cluster}  & \textbf{Technique 1} & \textbf{Technique 2} & \textbf{Technique 3} & \textbf{Technique 4} \\ \midrule
\multirow{2}{*}{TC1}                 
& \multirow{2}{*}{TS1, TS2, TS3, TS4}        
& \multirow{2}{*}{C1, C2, C3, C1} 
& \multirow{2}{*}{[C1, C2, C3]}
& \multirow{2}{*}{[1, 1, 1, 0, 0]}
& \multirow{2}{*}{[2, 1, 1, 0, 0]}
& \multirow{2}{*}{\shortstack[l]{[2, 1, 1, 0, 0] +  \\
{[}TC1 Name embedding{]} }} 
\\ \\
\multirow{2}{*}{TC2}                 
& \multirow{2}{*}{TS1, TS5, TS6, TS7, TS8} 
& \multirow{2}{*}{C1, C4, C5, C2, C5} 
& \multirow{2}{*}{[C1, C2, C4, C5]}
& \multirow{2}{*}{[1, 1, 0, 1, 1]}
& \multirow{2}{*}{[1, 1, 0, 1, 2]}
& \multirow{2}{*}{\shortstack[l]{[1, 1, 0, 1, 2] + \\ 
{[}TC2 Name embedding{]} }} 
\\ \\ \bottomrule
\end{tabular}
\end{table*}



In the example, there are two test cases (TC1 and TC2). TC1 contains four steps (TS1, TS2, TS3, TS4) and TC2 contains five steps (TS1, TS5, TS6, TS7, TS8). As we can see, only the TS1 step is shared between the test cases. In the test step cluster column, we can see the cluster ID to which each step belongs (TS1 belongs to the C1 cluster, TS2 belongs to the C2 cluster, and so on), where \textit{Cn} is the ID of the cluster $n$. Note that different steps (such as TS2 and TS7) might belong to the same cluster (C2). Next, we explain each proposed technique using this example.

\noindent\textbf{Technique 1: Test step cluster overlap.}
For this technique, we used only the identifiers of the test step clusters to represent test cases. For each test case, we gathered the unique list of cluster IDs that contain the test steps. For our running example, the TC1 test case is represented through the \textit{[C1, C2, C3]} vector, while TC2 is represented through the \textit{[C1, C2, C4, C5]} vector. Finally, we computed the pairwise similarity of any two test cases using a simple overlap metric, which indicates the proportion of overlap that test cases have in terms of test step cluster IDs, as shown below:

\begin{equation}
\text{Overlap} = \frac{ length((TCn) \cap (TCm)) }{ max(length(TCn),length(TCm)) }
\end{equation}

Where $TCn$ and $TCm$ correspond to the representations of the test cases $n$ and $m$ through the unique cluster IDs, respectively. Intuitively, test cases that have a large overlap of test step clusters (even if the test steps themselves are different) should be similar since test steps in the same cluster are (most of the time) similar. For our example, the length of TC1 is three (C1, C2, C3), the length of TC2 is four (C1, C2, C4, C5), and the length of the intersection between TC1 and TC2 is two (C1, C2). Therefore, the overlap between the TC1 and TC2 test cases is: \( \frac{2}{max(3,4)} = \frac{2}{4} = 0.5\) (50\%).

\begin{figure*}[]
\centering
\begin{subfigure}{0.235\textwidth}
\centering
    \includegraphics[width=0.95\linewidth]{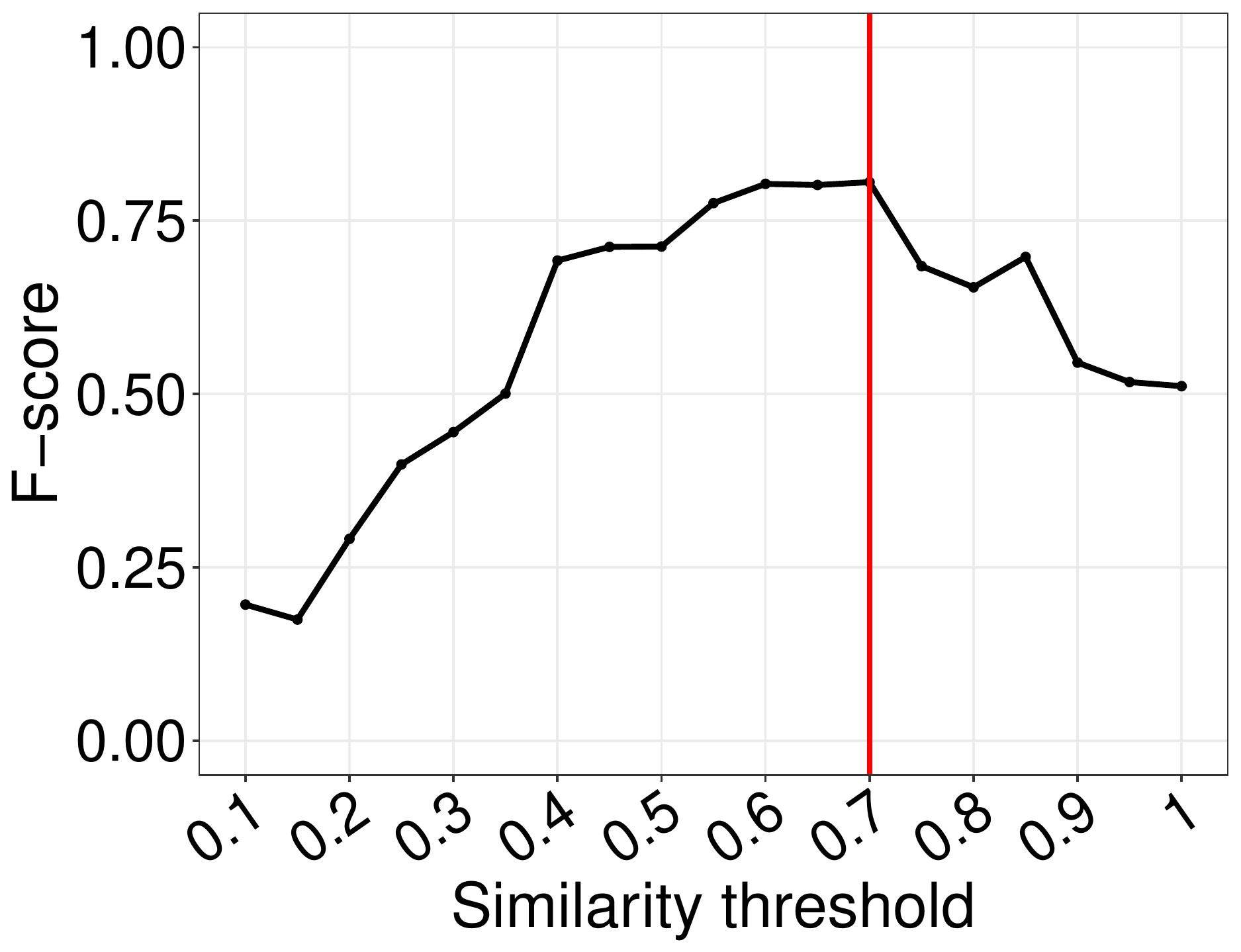}
    \ExerciseCaption{F-score for different similarity thresholds for Technique 1.}
    \label{fig:technique_1}
\end{subfigure} \hspace{0.01\textwidth} %
\begin{subfigure}{0.235\textwidth}
\centering
    \includegraphics[width=0.95\linewidth]{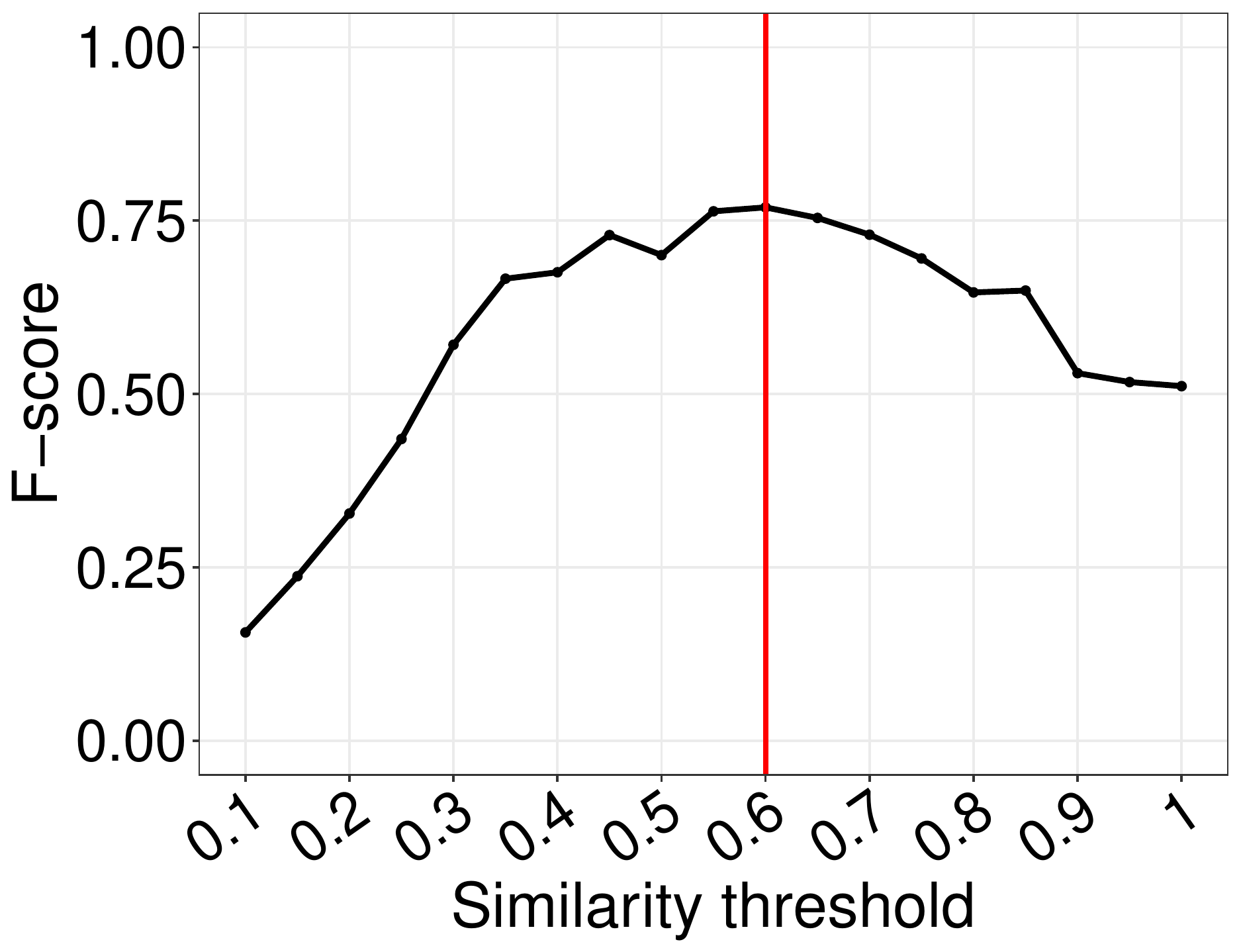}
    \ExerciseCaption{F-score for different similarity thresholds for Technique 2.}
    \label{fig:technique_2}
\end{subfigure}\hspace{0.01\textwidth} %
\begin{subfigure}{0.235\textwidth}
\centering
    \includegraphics[width=0.95\linewidth]{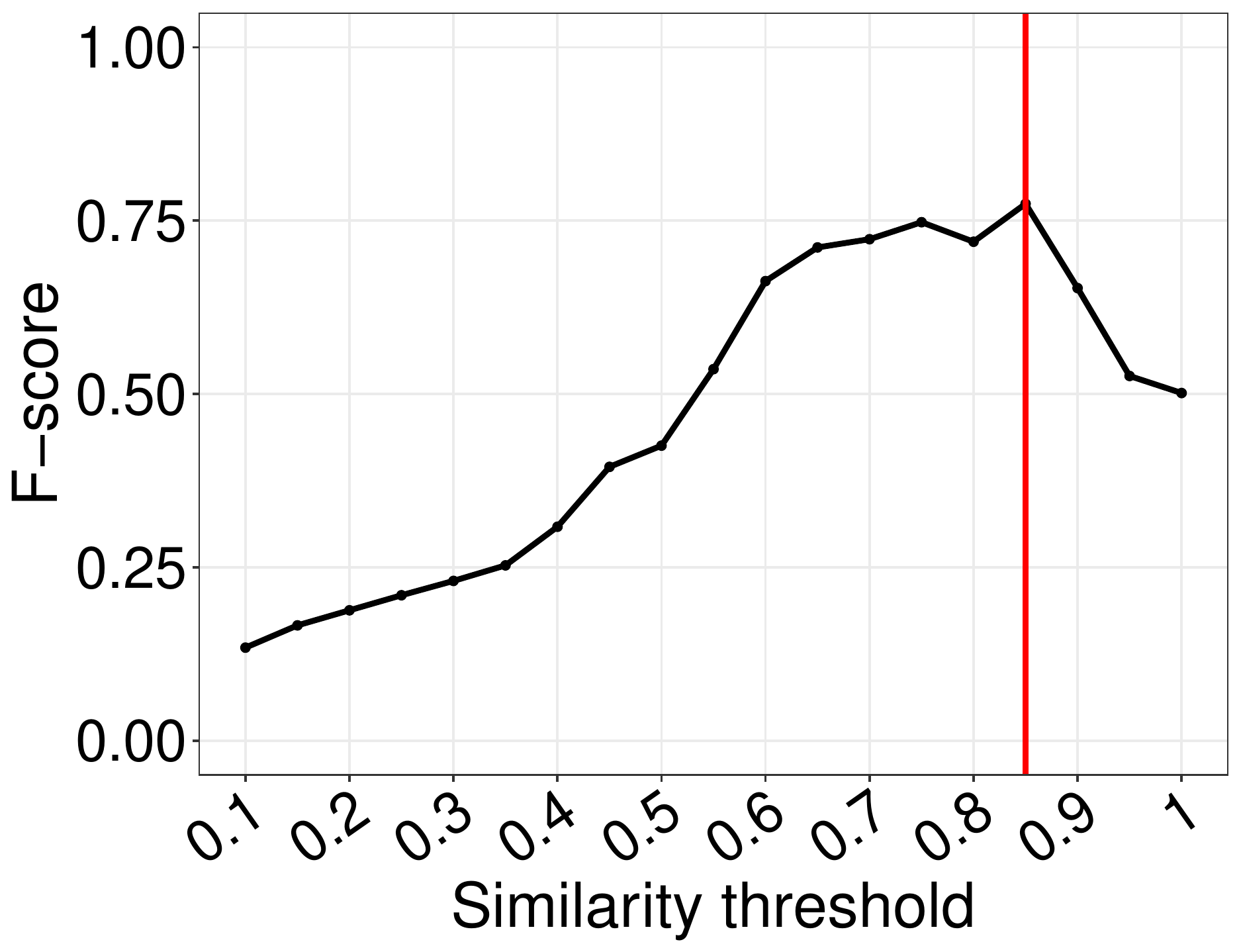}
    \ExerciseCaption{F-score for different similarity thresholds for Technique 3.}
    \label{fig:technique_3}
\end{subfigure}\hspace{0.01\textwidth} %
\begin{subfigure}{0.235\textwidth}
\centering
    \includegraphics[width=0.95\linewidth]{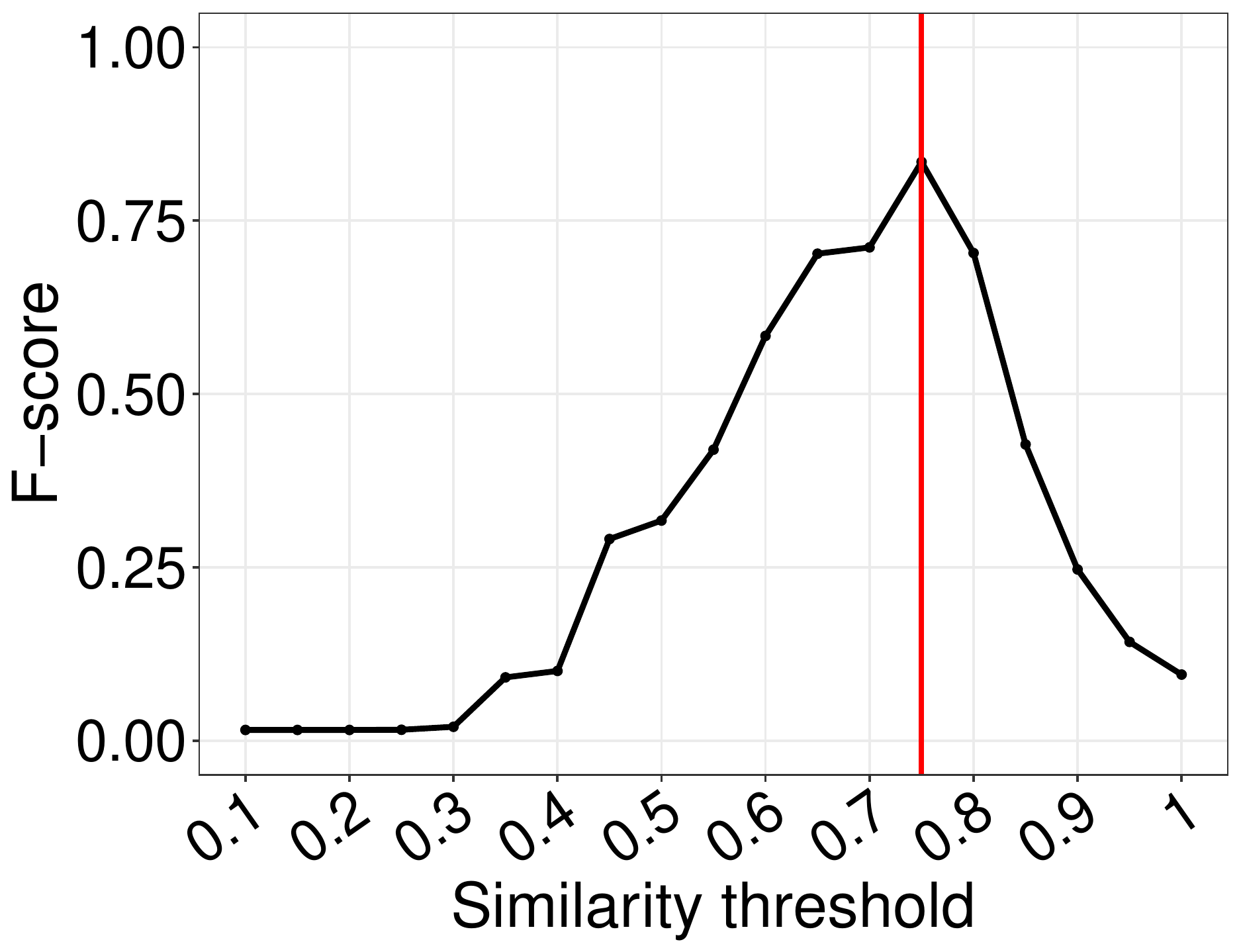}
    \ExerciseCaption{F-score for different similarity thresholds for Technique 4.}
    \label{fig:technique_4}
\end{subfigure}
\caption{F-score for different similarity thresholds for our four proposed techniques. The vertical red line indicates the threshold that maximizes the F-score.}
\label{fig:similarity_search_plots}
\end{figure*}

We used the computed overlap as the similarity metric to compare the test cases. Furthermore, in order to determine the optimal similarity threshold (i.e., with the optimal tradeoff between false positives and false negatives) to be used to identify similar test cases, we performed a search by varying the threshold from 0.1 (10\% of overlap) up to 1.0 (100\% of overlap). Figure~\ref{fig:technique_1} shows how the F-score changes with the similarity threshold (the optimal threshold is indicated with the vertical red line). As we can see, our search showed that the threshold that provides the maximum F-score is 0.70, which means that two test cases should be considered similar if their overlap metric is at least 70\%.



\noindent\textbf{Technique 2: Boolean representation of test cases.}
Similarly to Technique 1, for Technique 2 we used the test step clusters to represent test cases. However, instead of using the cluster IDs directly, we used a Boolean vector for each test case, in which we flagged the clusters that contain at least one test step of that case with a ``1''. Otherwise, we used ``0''. For our example, both test cases TC1 and TC2 are represented through a vector of length five because there are five different test step clusters in total (C1, C2, C3, C4, C5). TC1 is represented through the \textit{[1,1,1,0,0]} vector (because TC1 has steps that belong to the clusters C1, C2, and C3, but no step belongs to the clusters C4 and C5), while TC2 is represented through the \textit{[1,1,0,1,1]} vector. We built a matrix of dimension \textit{[\#test\_cases x \#test\_step\_clusters]}, where each row corresponds to a test case and each column corresponds to a test step cluster. Finally, we computed the pairwise similarity of any two test cases using the Jaccard index, as used in prior work~\cite{abreu2007accuracy, abreu2009practical} to calculate the similarity between Boolean vectors. Our search (see Figure~\ref{fig:technique_2}) shows that the optimal lower threshold for the Jaccard index is 0.60, which means that two test cases are similar if their Jaccard index is equal or larger than 0.60.


\noindent\textbf{Technique 3: Numeric representation of test cases.}
Using a Boolean vector to represent test cases might not be sufficient for situations where test cases have more than one step in a cluster. Therefore, we modified the previous technique so that, instead of representing test cases as a Boolean vector, we represent test cases as a numeric vector. This numeric vector corresponds to the number of test steps that the test case has in each cluster. For our example, TC1 is represented through the \textit{[2,1,1,0,0]} vector (because TC1 has two steps in the C1 cluster, one step in each of the C2 and C3 clusters, and no step in the C4 and C5 clusters). TC2 is represented through the \textit{[1,1,0,1,2]} vector.
We found that using a threshold of 0.85 (see Figure~\ref{fig:technique_3}) achieved the best performance in terms of F-score. This means that all the pairs of test cases that have a cosine similarity equal or larger than 0.85 are considered similar by Technique 3.

\begin{table*}[!t]
\centering
\caption{F-score of the test case similarity techniques along with the optimal similarity threshold. }
\label{tab:f_score_test_case_similarity}
\begin{tabularx}{1.5\columnwidth}{lp{5cm}rp{3.8cm}} \toprule
\textbf{Technique} & \textbf{Technique name}  &  \multicolumn{1}{r}{\textbf{F-score}}  & \multicolumn{1}{r}{\textbf{Optimal similarity threshold}} \\ \midrule
- & Baseline 1                   & 47.35 & \multicolumn{1}{r}{-}  \\ [1pt]
- & Baseline 2                   & 0.35   & \multicolumn{1}{r}{-} \\ [1pt]
Technique 1 &  Test step cluster overlap           & 80.54 & \multicolumn{1}{r}{0.70}    \\ [1pt]

Technique 2 &  Boolean representation of test cases            & 76.90 & \multicolumn{1}{r}{0.60}  \\ [1pt]

Technique 3 &  Numeric representation of test cases          & 77.42  & \multicolumn{1}{r}{0.85} \\ [1pt]

\multirow{2}{*}{\textbf{Technique 4}} & \textbf{Numeric representation of test cases with test case name embedding} & \multirow{2}{*}{\textbf{83.47}} &
\multicolumn{1}{r}{\multirow{2}{*}{\textbf{0.75}}}  \\ \bottomrule

\end{tabularx}
\end{table*}


\noindent\textbf{Technique 4: Numeric representation of test cases with test case name embedding.}
Our last technique is similar to Technique 3, but here we included information about the test case name as well. For our example, both TC1 and TC2 are represented through the same vectors as for Technique 3. In this technique, we combined the test step clusters with the test case name embedding. To obtain the embeddings for the name, we used the best-performing text embedding technique from the experiments for test step clustering (which is Word2Vec). Following a similar process as we did for the test step clustering, we computed the pairwise distance for any two test case name embeddings. We computed the weighted average between the test case name distance and the similarity score between the clusters of steps to obtain the final similarity score for the test cases. In order to determine the best weights for the test case name and test step cluster distances, and to determine the optimal threshold for the cosine similarity of the final distance (as we did for the Techniques 1, 2, and 3), we performed a search similarly to the previous techniques. We found that the test case name and the test step clusters must contribute equally (i.e., a weight of 50\% for each) to achieve the highest performance. Furthermore, as Figure~\ref{fig:technique_4} shows, using a threshold of 0.75 for this similarity score achieved the best performance, which means that all the pairs of test case pairs that have a cosine similarity equal or larger than 0.75 are considered similar by Technique 4. Note that, due to space constraints and for a better visualization, Figure~\ref{fig:technique_4} only displays how the F-score changes with the threshold already using the optimal weight of 50\%.

\noindent\textbf{Baseline.}
We compared the performance of the four approaches with two baselines. The first baseline considers two test cases to be similar if they have the exact same steps (regarding the text of the step). The second baseline considers two test cases to be similar if they have the same name.

\subsection{Evaluation metric.}
To evaluate our approaches for finding similar test cases, we used the manually built ground truth of similar test cases to compute the F-score. We followed the exact same process as we did previously for the test step clustering (Section~\ref{sec:evaluation_metric}).

\subsection{Findings}
\textbf{Clusters of similar test steps and test case name embeddings together can be used to identify similar test cases with a high performance.} Table~\ref{tab:f_score_test_case_similarity} presents the F-score of all the techniques that we evaluated along with the optimal similarity threshold.

We observe that Technique 4 (\textbf{Numeric representation of test cases with test case name embedding}) achieves the highest performance in terms of F-score (83.47\%), followed by Technique 1 (\textbf{Test step cluster overlap}), which achieves an F-score of 80.54\%. We also observe that, even though Technique 3 (\textbf{Numeric representation of test cases}) achieves a higher performance than Technique 2 (\textbf{Boolean representation of test cases}), the improvement is very small (0.52 in absolute percentage point). This indicates that using the number of test steps in each cluster (instead of just flagging whether the cluster contains a test step) slightly improves the performance of the test case similarity technique. Further incorporating the test case name information significantly improves the performance of the technique. Regarding the baselines, we observe that Baseline 2 (test cases with the same name) achieves an extremely low F-score (0.35\%), while Baseline 1 (test cases with the same steps) achieves a better, but still low F-score (47.35\%). Furthermore, all the experimented techniques perform considerably better than both baseline methods.

For the similarity threshold, we observe that each technique has a different optimal threshold. For example, even though we may expect that a higher overlap would achieve a higher performance, this does not hold true, as we see that the optimal overlap threshold (Technique 1) is 0.70. We also observe that the threshold for the highest-performing technique (Technique 4) is 0.75. Using this similarity threshold, Technique 4 found that 64.8\% of the test cases (which corresponds to 2,153 test cases) have at least one similar test case and there are, in total, 429 groups of similar test cases. On average, each cluster has two similar test cases, with a standard deviation of four.

\begin{table*}[t]
\centering
\caption{Examples of the four types of test case similarity. Differences between test cases' steps are highlighted in bold.}
\label{tab:test_case_similarity_type}
\begin{tabular}{lll} \toprule
\textbf{Similarity type}   & \textbf{Test case name}              
& \textbf{Test steps}       \\ \midrule
\multirow{4}{*}{\begin{tabular}[c]{@{}l@{}}(1) Same steps for different \\game assets\end{tabular}}
& \multirow{2}{*}{Check Hat - In Backpack}  
& 1. Verify item name                \\
&   & 2. Verify item icon                                            \\ \cmidrule{2-3}
& \multirow{2}{*}{Check Wand - In Backpack}       
& 1. Verify item name                 \\
&   & 2. Verify item icon                                            \\ \midrule
\multirow{4}{*}{\begin{tabular}[c]{@{}l@{}}(2) Slightly different steps \\for different game assets\end{tabular}}
& \multirow{2}{*}{Equip Hat}                        
& 1. Trigger equip functionality via \textbf{backpack hat} item slot      \\
&   & 2. Trigger unequip functionality via \textbf{backpack hat} item slot  
\\   \cmidrule{2-3}
& \multirow{2}{*}{Equip Wand}                       
& 1. Trigger equip functionality via \textbf{wand backpack} item slot     \\
&  & 2. Trigger unequip functionality via \textbf{wand backpack} item slot
\\ \midrule
\multirow{6}{*}{\begin{tabular}[c]{@{}l@{}}(3) Test cases with\\   additional/missing steps\end{tabular}}
& \multirow{4}{*}{Check Consumables (Water Resist)} 
& 1. Use in battle \\ 
&   & 2. Check battle bonus                                          \\
&   & 3.   \textbf{Check item card name}                                      \\
&  & 4.   \textbf{Check item card stats}                                     
\\ \cmidrule{2-3}
& \multirow{2}{*}{Check Food (Popcorn)}             
& 1. Use in battle                                               
\\
&   & 2. Check battle bonus                                          \\ \midrule
\multirow{2}{*}{\begin{tabular}[c]{@{}l@{}}(4) Redundant test cases\end{tabular}}
& Catch Firefly in Forest                           
& 1. \textbf{Catch firefly in forest}                                     
\\ \cmidrule{2-3}
& Firefly Forest - Catch Firefly                    
& 1. \textbf{Catch a  firefly}    
\\ \bottomrule
\end{tabular}
\end{table*}


Aiming at further understanding the output (groups of similar test cases) produced by our best technique (Technique 4), we manually inspected a representative sample of 100 of the obtained groups of similar test cases. We identified four main types of similar test cases, as shown in Table~\ref{tab:test_case_similarity_type}. While Type 1 corresponds to test cases with the same steps for different game assets, Type 2 regards test cases that have slightly different steps to indicate the asset being tested (e.g., \textbf{backpack hat} and \textbf{backpack wand}). Type 3 refers to test cases with a large overlap of steps but one of them has more/less steps, which might indicate extra (unnecessary) or missing steps. Finally, Type 4 regards redundant test cases, which are written differently and may have a different number of steps, but the testing objective is the same. The last type of similarity helps to identify test cases that might be completely removed from the test suite.




\section{Discussion}\label{sec:discussions}
In this section, we revisit the research questions and discuss the validation of our approach. 

\noindent\textbf{\rqone}\\
Our experiments demonstrate that we can identify similar test steps with a high performance in terms of F-score. We showed that an ensemble approach using a combination (majority voting) of different techniques (five text embedding techniques with two similarity metrics and two clustering algorithms) achieves the highest performance. Such ensemble approach has a large computational cost as it requires the execution of several different techniques. However, we showed that using a single technique (such as Word2Vec or TF-IDF) can also provide a high performance while being less computationally expensive.

\noindent\textbf{\rqtwo}\\
Our experiments demonstrate that we can use the clusters of similar test steps identified in the first part of the study to represent test cases and identify the similar ones. More specifically, representing test cases through a vector that captures the number of test steps in each cluster boosts the similarity technique performance. Furthermore, we showed that combining the clusters of similar test steps with the embedding of the test case name achieves an even higher performance. Our experiments showed that the optimal weight (for our data) for the clusters and the test case names is 50\%. In addition, we can use a threshold of 0.75 for the similarity score to decide whether two test cases are similar.

\noindent\textbf{Validation with developers.}
To validate the results of our approach, we did an informal interview with a QA expert at Prodigy Education to discuss whether our results are valid and how they can be used in practice and improve the testing process. We selected a purposive sample~\cite{tongco2007purposive,guarte2006estimation,etikan2016comparison} to explicitly select test cases that cover the different types of similar test cases that we identified.

Overall, the expert validated the different types of test case similarity that we identified and mentioned that our approach can help the QA to improve the quality of the test cases. More specifically, the QA expert pointed out five practical usages of our approach, as we explain next. First, our approach can be used to identify redundant test cases. Such cases occur when test cases are described differently (e.g., because they were written by different professionals) but test exactly the same aspect/asset of the game. The second usage regards the reuse of existing test cases when creating new ones for new features of the game. In this case, existing test cases can either be fully or partially (e.g., a few test steps) reused. By reusing test cases, the overall quality of the test suite improves with more consistent and homogeneous descriptions in terms of terminology. Furthermore, reusing test cases reduces the manual effort and time required for designing and creating new test cases. The third application is to use our approach to identify legacy test cases, such as test cases that were created for a temporary feature and were kept in the test case base, even though the feature does not exist anymore. Such legacy test cases have a high impact in the (already time-consuming) manual testing effort. The fourth usage is to identify test cases with missing steps. A few test case samples that we discussed with the expert were indeed groups of similar test cases which perform the same task, but some of the cases had less steps than what is actually performed by a tester. We further investigated those cases with the QA expert and found out that the missing steps were scattered across the test suite (in different cases) and should be merged with the steps of the main test case. Finally, the fifth application is to identify test cases which are redundant but one of the cases has additional steps. This occurs when new test cases are created based on existing ones, but some steps are added for clarification purposes and the older test case is not removed from the test suite.

\section{Threats to Validity}\label{sec:threats}
\textbf{External validity} relates to the generalizability of our findings. One threat is regarding the methods that we used for text embedding, text similarity and clustering. Although we used five different text embedding techniques, three similarity metrics and two clustering algorithms, different results might be achieved with other methods. Future studies should further investigate additional methods for text embedding, text similarity and clustering. Another threat is that our findings rely on the test case descriptions of an educational game company. Test cases of organizations from different domains might be different (e.g., in terms of the used terminology and grammar complexity and structure) and might affect the results. Finally, our thresholds for optimal values (such as the number of clusters and the similarity metric) likely do not apply to other systems. However, our method for conducting the search for these values is generalizable.

\textbf{Internal validity} concerns the bias and errors due to the experimental design. One threat is related to the manual analysis of the samples of test steps (to build the ground truth for the test step clustering) and test cases (to build the ground truth for the test case similarity). The manual analysis is subject to error and bias because of human factors. Furthermore, despite selecting statistically representative samples, the characteristics of the whole population may not be represented in those samples.




\section{Conclusion}\label{sec:conclusion}
Test cases written in natural language are often defined by different people who may use different terminology to refer to the same concept. As a result, many similar or redundant test cases may exist in the test suite, which increases the manual testing effort and the usage of development resources. Since manually identifying redundant test cases is a time-consuming task, an automated technique is necessary.

In this paper, we propose an approach to identify similar test cases specified in natural language. First, we evaluated different text embedding techniques, similarity metrics, and clustering algorithms to identify clusters of similar test steps (which compose test cases). We then leveraged the identified test step clusters together with the test case name to identify similar test cases. To evaluate the approach, we used test cases from an educational game company. We manually built a ground truth of similar test steps and test cases and computed the F-score metric. The approach evaluation shows that similar test steps can be identified with a high performance (an F-score of 87.39\%) using an ensemble approach which consists of different NLP techniques. We can also achieve a similar performance (an F-score of 86.99\%) using a single technique (Word2Vec). Furthermore, we identified similar test cases with a high performance (an F-score of 83.47\%) using clusters of similar test steps combined with the similarity between test case names.

In this work, we show how we can identify similar test cases based only on their description in natural language with an unsupervised approach, which requires no labelled data nor human supervision. As indicated in an informal interview with a QA engineer, our approach has several usages in practice, such as supporting QA and developers to identify and remove redundant and legacy test cases from the test suite. Furthermore, existing groups of similar test cases can be leveraged to create new test cases and help to maintain a more consistent and homogeneous terminology across the test suite.

\section*{Acknowledgments} 
The research reported in this article has been supported by Prodigy Education and the Natural Sciences and Engineering Research Council of Canada under the Alliance Grant project ALLRP 550309.

\ifCLASSOPTIONcaptionsoff
  \newpage
\fi

\footnotesize

\bibliographystyle{IEEEtranSN}
\bibliography{IEEEabrv, bibliography}

\end{document}